\documentclass[a4paper,UKenglish,cleveref, autoref, thm-restate]{lipics-v2021}

\title{Convergence of the number of period sets in strings} 

\titlerunning{Convergence of the number of period sets in strings} 

\author{Eric Rivals}{LIRMM, Université Montpellier, CNRS, Montpellier, France \and \url{https://www.lirmm.fr}}{rivals@lirmm.fr}{https://orcid.org/0000-0003-3791-3973}{}

\author{Michelle Sweering}{CWI, Amsterdam, The Netherlands \and \url{https://michellesweering.github.io/}}{michelle.sweering@cwi.nl}{https://orcid.org/0000-0003-1200-6015}{} 

\author{Pengfei Wang}{LIRMM, Université Montpellier, CNRS, Montpellier, France \and \url{https://www.lirmm.fr/~rivals/authors/pengfei-wang/}} {pengfei.wang@lirmm.fr}{https://orcid.org/0000-0001-8172-5270}{}

\authorrunning{E. Rivals, M. Sweering, and P. Wang} 

\Copyright{Eric Rivals, Michelle Sweering, and Pengfei Wang} 



\category{} 

\supplement{}

\funding{E. Rivals and P. Wang are supported by the European Union’s Horizon 2020 research and innovation programme under the Marie Skłodowska-Curie grant agreement No 956229. M.~Sweering is supported by the Netherlands Organisation for Scientific Research (NWO) through Gravitation-grant NETWORKS-024.002.003.}

\nolinenumbers 

\usepackage[table]{xcolor}
\def\dd{\mathinner{.\,.}}

\begin{document}
\pagestyle{empty}
\maketitle

\begin{abstract}
 Consider words of length $n$. The set of all periods of a word of length $n$ is a subset of $\{0,1,2,\ldots,n-1\}$. However, any subset of $\{0,1,2,\ldots,n-1\}$ is not necessarily a valid set of periods. In a seminal paper in 1981, Guibas and Odlyzko proposed to encode the set of periods of a word into an $n$ long binary string, called an autocorrelation, where a one at position $i$ denotes the period $i$. They considered the question of recognizing a valid period set, and also studied the number of valid period sets for strings of length $n$, denoted $\kappa_n$. They conjectured that $\ln(\kappa_n)$ asymptotically converges to a constant times $\ln^2(n)$. Although improved lower bounds for $\ln(\kappa_n)/\ln^2(n)$ were proposed in 2001, the question of a tight upper bound has remained open since Guibas and Odlyzko’s paper. Here, we exhibit an upper bound for this fraction, which implies its convergence and closes this longstanding conjecture. Moreover, we extend our result to find similar bounds for the number of correlations: a generalization of autocorrelations which encodes the overlaps between two strings.
\end{abstract}
\newpage
\section{Introduction}
A linear word can overlap itself if one of its prefixes is equal to one of its suffixes. The corresponding prefix (or suffix) is called a border and the shift needed to match the prefix to the suffix is called a period.  The dual notions of period and border are critical concepts in word combinatorics:  important definitions such as periodic and primitive words, or the normal form of a word rely on them. These concepts play a role in key results of the field like the  Critical Factorization Theorem~\cite{Lothaire-COW-97}.
In computer science, in the field of string algorithms (a.k.a. stringology),  pattern matching algorithms heavily exploit borders/periods to optimize the search of occurrences of a word in a text~\cite{Smyth-book-03}.  For clarity, note that the terms \emph{word} and \emph{string} both mean a sequence of letters taken from an alphabet.  These notions also play a role in statistics.  The set of periods of a word controls how two occurrences of the same word can overlap in a text. Hence, the set of periods (or autocorrelation) is a key variable to study the statistics of word occurrences in random texts (waiting time, distance between successive occurrences, etc.)~\cite{Robin-etal-dna-book}. The notion of autocorrelation has been extended to describe how two distinct words can have overlapping occurrences in the same text. This has been used for instance to study the number of missing words in random texts~\cite{rahmann_cpc_2003} or to design procedures for testing pseudo-random number generators~\cite{PER:WHI:1995}. 

\bigskip
An autocorrelation is a binary vector representing the set of periods of a word. The concept of autocorrelation was introduced by Guibas and Odlyzko in~\cite{Guibas}. They gave a characterization of autocorrelations and proved the following bounds on $\kappa_n$ - the cardinality of the set $\Gamma_n$ of autocorrelations of words of length $n$.
$$\frac{1}{2\ln (2)} +o(1)  \leq \frac{\ln(\kappa_n)}{\ln^2(n)} \leq \frac{1}{2\ln (3/2)} +o(1)$$ 
They conjectured that $\ln(\kappa_n)$ is asymptotic to a constant times $\ln^2(n)$. Rivals and Rahmann~\cite{Rivals}, later on studied the combinatorial structure of the set of autocorrelations $\Gamma_n$, and improved the lower bound on $\kappa_n$ as follows:
$$\frac{\ln(\kappa_n)}{\ln^2 (n)} \geq \frac{1}{2\ln (2)}\left(1 - \frac{\ln(\ln(n))}{\ln(n)}\right)^2 + \frac{0.4139}{\ln(n)} - \frac{1.47123\ln(\ln(n))}{\ln^2 (n)} +O\left(\frac{1}{\ln^2 (n)}\right).$$
However, to date, no one has focused on improving the upper bound on $\kappa_n$. In this work, we apply the notion of irreducible period sets introduced by Rivals and Rahmann~\cite{RivalsRahmann-ICALP01,Rivals} to prove that
$$ \frac{\ln(\kappa_n)}{\ln^2(n)}  \leq \frac{1}{2 \ln(2)} + \frac{3}{2\ln(n)} \quad \forall n \in \mathbb{N}_{\geq 2}.$$
Together with known asymptotic lower bounds~\cite{Rivals}, we find that
$$\frac{\ln\kappa_n}{\ln^2(n)} \rightarrow \frac{1}{2\ln(2)} \quad \text{as} \quad n \rightarrow \infty,$$
thus resolving the conjecture of Guibas and Odlyzko.

In their paper about autocorrelations~\cite{Guibas}, Guibas and Odlyzko also introduced the notion of correlation between words. For two words $u$ and $v$, the \emph{correlation} of $u$ over $v$ is a binary vector indicating all overlaps between suffixes of $u$ and prefixes of $v$. In particular, an autocorrelation is the correlation of a word with itself.
We show that the number of correlations between two words of length $n$, denoted by $\delta_n$, has the same asymptotic convergence behaviour as the number of autocorrelations of words of length $n$, that is
$$\frac{\ln\delta_n}{\ln^2(n)} \rightarrow \frac{1}{2\ln(2)} \quad \text{as} \quad n \rightarrow \infty.$$

\subsection{Related works}
\label{sec:org508bc2f}
Apart from previously cited articles that deal with the combinatorics of period sets, some related works exist in the literature.

For instance, the question of the average period of a random word has been investigated in \cite{holub_shallit_random_words}.  Clearly, the number of periods of a word of length \(n\) lies between one and \(n\).  A recent work exhibits an upper bound on the number of periods of a word as a function of its  \emph{initial critical exponent} -- a characteristic of the word related to its degree of periodicity  \cite{gabric_etal_inequality_periods_2021}, but this has not been used to bound the number of period sets.
Last, the combinatorics of period sets has also been investigated in depth for a generalization of the notion of words, called \emph{partial words}~\cite{Blanchet-etal-STACS-2007}.  In partial words, some positions may contain a \emph{don't care} symbol, which removes some constraints of equality between positions. To study the combinatorics of period sets, Blanchet-Sadri \emph{et al.} defined weak and strong periods, and proved several important theorems \cite{Blanchet-Duncan-JCTA-2005}, including lower and upper bounds on the number of binary and ternary autocorrelations \cite{Blanchet-etal-STACS-2007,Blanchet-etal-JCTA-2010}. However, the cardinality of the family of period sets differs between normal words and partial words, and a tight  upper bound for normal words cannot be deduced from that for partial words.
Several works investigate sets of words with constraints (either absence or presence) on their mutual overlaps: mutually bordered (overlapping) pairs of words are studied in \cite{Gabric}, while methods for constructing a set of mutually unbordered words (also called cross-bifix-free words) are provided in \cite{Bilotta},   
\cite{Bajic}, \cite{bilotta2017}. 

\section{Preliminaries}
A string $u = u[0 \dd n-1] \in \Sigma^n$ is a sequence of $n$ letters over a finite alphabet $\Sigma$. For any $0\leq i \leq j \leq n - 1$, we denote the substring starting at position $i$ and ending at position $j$ with $u[i \dd j]$. In particular, $u[0 \dd j]$ denotes a prefix and $u[i \dd n-1]$ a suffix of $u$. Throughout this paper, all our strings and vectors will be zero-indexed.

\subsection{Periodicity}
In this subsection, we define the concepts of period, period set, basic period, and autocorrelation, and then review some useful results.
For the sake of self-containment, we provide in Appendix~\ref{sec:omitted} the proofs for all lemmas of this subsection.


\begin{definition}[Period]\label{def:period1}
String $u = u[0 \dd n-1]$ has a period $p \in \{1, \ldots, n-1\}$ if and only if for any $0 \leq i,j \leq n - 1$ such that $i \equiv j \mod p$, we have $u[i] = u[j]$. Moreover, we consider that $p = 0$ is a period of any string of length $n$.\end{definition}
An equivalent definition is the following.
\begin{definition}[Period]\label{def:period2}
The string $u = u[0 \dd n-1]$ has period $p \in \{0, 1, \ldots, n-1\}$ if and only if $u[0\dd n-p-1] = u[p\dd n-1]$, i.e. for all $0 \leq i \leq n-p-1$, we have $u[i] = u[i+p]$.
\end{definition}

The smallest non-zero period of $u$ is called its \emph{basic period}. The \emph{period set} of a string $u$ is the set of all its periods and is denoted by $P(u)$. 
We will now list some useful properties about periods, which we will need later on. Their proofs can be found in~\cite{Guibas,Lothaire-COW-97} and in Appendix~\ref{sec:omitted}.

\begin{lemma}\label{lem:multiply}
Let $p$ be a period of $u \in \Sigma^n$ and $k \in \mathbb{Z}_{\geq 0}$ such that $kp < n$. Then $kp$ is also a period of $u$.
\end{lemma}
\begin{lemma}\label{lem:add}
Let $p$ be a period of $u \in \Sigma^n$ and $q$ a period of the suffix $w = u[p\dd n-1]$. Then $(p + q)$ is a period of $u$. Moreover, $(p + kq)$ is also a period of $u$ for all $k \in \mathbb{Z}_{\geq 0}$ with $p + kq < n$.
\end{lemma}
\begin{lemma}\label{lem:subtract}
Let $p, q$ be periods of $u \in \Sigma^n$ with $0 \leq q \leq p$. Then the prefix and the suffix of length $(n-q)$ have the period $(p-q)$.
\end{lemma}

\begin{lemma}\label{lem:divide}
Suppose $p$ is a period of $u \in \Sigma^n$ and there exists a substring $v$ of $u$ of length at least $p$ and with period $r$, where $r|p$. Then $r$ is also a period of $u$.
\end{lemma}

We will also use the famous Fine and Wilf theorem~\cite{FineWilf}, a.k.a. the periodicity lemma, for which a short proof was provided by Halava and colleagues~\cite{HALAVA2000298}.
\begin{theorem}[Fine and Wilf]\label{thm:gcd}
Let $p, q$ be periods of $u \in \Sigma^n$. If $n \geq p + q - \gcd(p,q)$, then $\gcd(p,q)$ is a period of $u$.
\end{theorem}

\subsection{Autocorrelation}
We now give a formal definition of an autocorrelation.
\begin{definition}[Autocorrelation]\label{def:autocorrelation}
For every string $u \in \Sigma^n$, its autocorrelation is the string $s \in \{0, 1\}^n$ such that
$$s[i] = \begin{cases}1 & \text{if } i \text{ is a period of } u\\
0 & \text{otherwise}
\end{cases} \quad \forall i \in \{0, \ldots, n - 1\}.
$$
\end{definition}
To illustrate this concept, consider the following example (detailed in Table~\ref{tab:auto}).
\begin{example} We consider the word $u = \mathtt{abbaabba}$ of length $8$. Its period set is $P(u) = \{0,4,7\}$, its basic period is 4 and its autocorrelation is $s = 10001001$. See Table~\ref{tab:auto}.
\end{example}
\begin{table}
\begin{center}
\begin{tabular}{>{\columncolor[gray]{.8}}cccccccccccccccc>{\columncolor[gray]{.8}}c} 
  \rowcolor[gray]{.8}
  pos. &\color{blue}{0} & 1 & 2 & 3 & \color{blue}{4} & 5 & 6 & \color{blue}{7} & 8 & 9 & 10 & 11 & 12 & 13 & 14 & 
  \\ 
  $u$ &\texttt{a} & \texttt{b} & \texttt{b} & \texttt{a} & \texttt{a} & \texttt{b} & \texttt{b} & \texttt{a} & - & - & - & - & - & - & - &$s$
  \\ \hline
  $u$ &\color{blue}{\texttt{a}} & \color{blue}{\texttt{b}} & \color{blue}{\texttt{b}} & \color{blue}{\texttt{a}} & \color{blue}{\texttt{a}} & \color{blue}{\texttt{b}} & \color{blue}{\texttt{b}} & \color{blue}{\texttt{a}} & - & - & - & - & - & - & - & \color{blue}{1}
  \\ 
  & - & \texttt{a} & \texttt{b} & \texttt{b} & \texttt{a} & \texttt{a} & \texttt{b} & \texttt{b} & \texttt{a} & - & - & - & - & - & - & 0
  \\
  & - & - & \texttt{a} & \texttt{b} & \texttt{b} & \texttt{a} & \texttt{a} & \texttt{b} & \texttt{b} & \texttt{a} & - & - & - & - & - & 0
  \\ 
  & - & - & - & \texttt{a} & \texttt{b} & \texttt{b} & \texttt{a} & \texttt{a} & \texttt{b} & \texttt{b} & \texttt{a} & - & - & - & - & 0
  \\ 
  & - & - & - & - & \color{blue}{\texttt{a}} & \color{blue}{\texttt{b}} & \color{blue}{\texttt{b}} & \color{blue}{\texttt{a}} & \color{blue}{\texttt{a}} & \color{blue}{\texttt{b}} & \color{blue}{\texttt{b}} & \color{blue}{\texttt{a}} & - & - & - & \color{blue}{1}
  \\ 
  & - & - & - & - & - & \texttt{a} & \texttt{b} & \texttt{b} & \texttt{a} & \texttt{a} & \texttt{b} & \texttt{b} & \texttt{a} & - & - & 0
  \\ 
  & - & - & - & - & - & - & \texttt{a} & \texttt{b} & \texttt{b} & \texttt{a} & \texttt{a} & \texttt{b} & \texttt{b} & \texttt{a} & - & 0
  \\
  & - & - & - & - & - & - & - & \color{blue}{\texttt{a}} & \color{blue}{\texttt{b}} & \color{blue}{\texttt{b}} & \color{blue}{\texttt{a}} & \color{blue}{\texttt{a}} & \color{blue}{\texttt{b}} & \color{blue}{\texttt{b}} & \color{blue}{\texttt{a}} & \color{blue}{1}
  \\
  \hline
\end{tabular}
\end{center}
\caption{This table illustrates the set of period and the autocorrelation of the word $u = \mathtt{abbaabba}$ of length $8$. A first copy of the word $u$ is shown on the second line. Another copy of $u$ is displayed on (each) line $(3+i)$ shifted by $i$ positions to the right, with $i$ going from $0$ to $7$. If the suffix of the copy of $u$ matches the prefix of the first copy $u$ on line $2$, then $i$ is a period, and both the line and the corresponding position/shift (on the first line) are colored in blue.  The last column contains the autocorrelation of $u$, with \texttt{1} bits corresponding to periods colored in blue.}\label{tab:auto}
\end{table}

Guibas and Odlyzko~\cite{Guibas} show that \emph{any alphabet of size at least two will give rise to the same set of correlations} (Corollary 5.1). Autocorrelations have many other useful properties~\cite{Guibas,Rivals}. The most significant one for our work is the following.
\begin{lemma}[Lemma 3.1~\cite{Guibas} and Theorem 1.3~\cite{Rivals}]\label{lem:closed}
If $s \in \{0, 1\}^n$ is an autocorrelation and $s[i] = 1$, then $s[i\dd n-1]$ is the autocorrelation of $u[i\dd n-1]$
\end{lemma}
\begin{proof}
Note that $s[i] = 1$ means: $i$ is a period of $u$. 
Suppose $s[i + j] = 1$. Then $i + j$ is a period of $u$. Thus $u[i\dd n-1]$ has period $(i + j) - i = j$ by Lemma~\ref{lem:subtract}.
Conversely, suppose $u[i\dd n-1]$ has period $(i + j) - i = j$. Then $i + j$ is a period of $u$ by Lemma~\ref{lem:add}. Thus $s[i + j] = 1$.
Combining these results, we find that $s[i+j] = 1$ if and only of $j$ is a period of $u[i\dd n-1]$, and equivalently $s[i\dd n - 1]$ is the autocorrelation of $u[i\dd n-1]$.
\end{proof}

\subsection{Irreducible Period Set }
To prove the upper bound on the number of autocorrelations, we use the notion of irreducible period sets as introduced by Rivals and Rahmann~\cite{Rivals}. An irreducible period set is the minimum subset of a period set that determines the period set using the Forward Propagation Rule. Before formally introducing the irreducible period set, we will first explain what forward propagation is.

\begin{lemma}[Forward Propagation Rule] Let $p \leq q$ be periods of a string $u$ of length $n$ and let $k \in \mathbb{Z}_{\geq 0}$ such that $p + k(q - p) < n$. Then $p + k(q - p)$ is a period of $u[0\dd n - 1]$.
\end{lemma}
\begin{proof}
It follows from Lemma~\ref{lem:subtract} that $u[p\dd n - 1]$ has period $q - p$. Applying Lemma~\ref{lem:add} we find that $u[0\dd n - 1]$ has period $p + k(q - p)$ for all $k \in \mathbb{Z}_{\geq 0}$. 
\end{proof}
The forward closure $FC_n(S)$ of a set $S \subseteq \{0,\ldots,n-1\}$ (not necessarily a period set, typically a subset of one) is the closure of $S$ under the forward propagation rule.
\begin{definition}[Forward Closure]\label{def:FC}
Let $S \subseteq \{0,\ldots, n-1\}$. Its forward closure $FC_n(S)$ is the minimum superset of $S$ such that for any $p, q \in FC_n(S)$ and $k \geq 0$ with $p \leq q$ and $p + k(q - p) < n$, we have $$p + k(q - p) \in FC_n(S).$$
\end{definition}
We can now define the irreducible period set.
\begin{definition}[Irreducible Period Set]\label{def:IPS}
Let $P$ be the period set of a string $u \in \Sigma^n$. An irreducible period set of $P$ is a minimal subset $R(P) \subseteq P$ with forward closure $P$. 
\end{definition}
Observe that there exists an irreducible period set for any period set $P$, because $FC_n(P) = P$ by the forward propagation rule. We will now give a useful characterization of an irreducible period set as the set of periods which are not in the forward closure of the set of all smaller periods. Consequently, every period set has exactly one irreducible period set, whose elements we will call irreducible periods.

Recall that for a given length $n$, we denote the set of all period sets by $\Gamma_n$. Formally stated $\Gamma_n$ is defined as:
$$\Gamma_n = \{ S \subseteq \{0, 1, \ldots, n-1\} : \exists u \in \Sigma^n \text{ such that } P(u) = S \}.$$
As in \cite{Rivals}, for a given length $n$,  we denote the set of all irreducible period sets by $\Lambda_n$:
$$\Lambda_n = \{ T \subseteq \{0, 1, \ldots, n-1\} : \exists u \in \Sigma^n \text{ such that } R(P(u)) = T \}.$$
The bijective relation between period sets and irreducible period sets implies that $|\Gamma_n| = |\Lambda_n|$.
\begin{lemma}
Let $P$ be the period set of a string $u \in \Sigma^n$ and $R(P)$ an irreducible period set of $P$. Then
$$R(P) = \left\{q \in P \mid q \not\in FC_n(P \cap [0,q-1])\right\}.$$
\end{lemma}
\begin{proof}
Let $p \in P$. We will prove the two alternative cases separately:
\begin{enumerate}[(a)]
    \item $p \not\in \left\{q \in P \mid q \not\in FC_n(P \cap [0,q-1])\right\} \implies p \not\in R(P)$ and
    \item $p \in \left\{q \in P \mid q \not\in FC_n(P \cap [0,q-1])\right\} \implies p \in R(P)$.
\end{enumerate}


\begin{enumerate}[(a)]
\item Suppose $p \not\in \left\{q \in P \mid q \not\in FC_n(P \cap [0,q-1])\right\}$, or equivalently $p \in FC_n(P \cap [0,p-1])$. Then 
\begin{align*}
    p \in FC_n(P \cap [0,p-1]) &= FC_n(FC_n(R(P)) \cap [0,p-1])\\
    &\subseteq FC_n(FC_n(R(P) \cap [0,p-1]))\\
    &= FC_n(R(P) \cap [0,p-1])\\
    &\subseteq FC_n(R(P) \setminus \{p\}).
\end{align*}
It follows that $FC_n(R(P) \setminus \{p\}) = FC_n(R(P))$. By minimality of irreducible period sets, we have $p \not\in R(P)$.
\item Suppose on the other hand that $p \not\in FC_n(P \cap [0,p-1])$. 
Then $p \not\in FC_n(P \setminus \{p\})$ either.
As $$FC_n(P \setminus \{p\}) \supseteq FC_n(R(P) \setminus \{p\}),$$
then $p \not\in FC_n(R(P) \setminus \{p\})$.

However, as $p \in P$ and $P = FC_n(R(P))$, it follows that $p \in R(P)$.
\end{enumerate}
\end{proof}

\section{Asymptotic convergence of \texorpdfstring{$\kappa_n$}{Lg}}

In this section, we present a new upper bound on $\kappa_n$, the number of distinct autocorrelations of strings of length $n$. Moreover, we shall prove that $\ln(\kappa_n)$ asymptotically converges to $c \cdot \ln^2(n)$, where $c = \tfrac{1}{2\ln(2)}$.

\begin{theorem}[Upper bound on $\kappa_n$]\label{thm:upp}
For all $n \in \mathbb{N}_{\geq 2}$ we have 
$$\frac{\ln(\kappa_n)}{\ln^2(n)}  \leq \frac{1}{2 \ln(2)} + \frac{3}{2\ln(n)}.$$
\end{theorem}
\begin{proof}
    To prove this theorem, we need several lemmas.
\begin{lemma}\label{lem:periodbound}
Let $u \in \Sigma^n$ with autocorrelation $s$, period set $P$, and irreducible period set $R(P) = \{0 = a_0 < \ldots < a_i < \ldots < a_k < n\}$. Then for all $0 \leq i \leq k$, there exists $q_i \in \{1,\ldots,n - a_i\}$ such that
\begin{enumerate}
    \item  $q_i \leq n / 2^i$, and
    \item  $a_i + q_i = n$ or $a_i + q_i$ is in the forward closure of $\{a_0, \ldots,
    a_i\}$.
\end{enumerate} 
\end{lemma}

\begin{proof} We will prove this by induction.
\paragraph*{Basis} 
By picking $q_0 = n \in \{1, \ldots, n - a_0\}$, we satisfy both $q_0 \leq n/2^0$ and $a_0 + q_0 = n$.

\paragraph*{Hypothesis} For some $0 \leq i < k$, there exists a $q_i \in \{1,\ldots,n - a_i\}$ such that
\begin{enumerate}
    \item  $q_i \leq n / 2^i$, and
    \item  $a_i + q_i = n$ or $a_i + q_i$ is in the forward closure of $\{a_0, \ldots,
    a_i\}$.
\end{enumerate}
\paragraph*{Step} We first note that if $n - a_{i+1} \leq n / 2^{i+1}$, then we can pick $q_{i+1} = n - a_{i+1}$. Suppose on the other hand that $n - a_{i+1} > n / 2^{i+1}$. We distinguish two cases.
\begin{itemize}
    \item If $a_i +q_i = n$, then 
     \begin{align*}
         a_{i+1} - a_i &= (n - a_i) - (n - a_{i+1})
         \\
         & < n/2^i - n/2^{i+1} 
         \\
         &= n / 2^{i+1} 
         \\
         &< n - a_{i+1}.
     \end{align*}
    Thus, we can pick $q_{i+1} = a_{i+1}-a_i \in \{1,\ldots,n - a_{i+1}\}$, since
\begin{enumerate}
    \item it satisfies $q_{i+1} \leq n / 2^{i+1}$ and
    \item $a_{i+1} \, +\, q_{i+1} = a_i \, +\, 2(a_{i+1} - a_{i})$ is in the forward closure of $\{a_0, \ldots,
    a_{i+1}\}$.
\end{enumerate}
    \item If $a_i + q_i$ is in the forward closure of $\{a_0, \ldots,
    a_i\}$, then 
    $$a_i + \lambda q_i = a_i + \lambda (a_i+q_i - a_i)$$ is in the forward closure of $\{a_0, \ldots,
    a_i\}$ for all integers $0 \leq \lambda \leq (n - 1 - a_i) / q_i$. Since $a_{i+1}$ is an irreducible period, there must exist an integer $\lambda_0 \in [0,(n - 1 - a_i) / q_i]$ such that
    $$a_i + \lambda_0 q_i < a_{i+1} < a_i + (\lambda_0 + 1)q_i.$$
    In other words, $a_{i+1}$ is comprised between two successive, non-irreducible periods generated from $a_i$ and $q_i$ using the FPR (or $n \leq a_i + (\lambda_0 + 1)q_i$).
    We pick 
    \begin{align*}
        q_{i+1} &= \min(a_{i+1} - (a_i + \lambda_0 q_i), (a_i + (\lambda_0 + 1)q_i) - a_{i+1}, n - a_{i+1})\\
        \intertext{ and note that }
        q_{i+1} &\leq \frac{a_{i+1} - (a_i + \lambda_0 q_i) + (a_i + (\lambda_0 + 1)q_i) - a_{i+1}}{2} 
        \\
        &= q_i / 2 
        \\
        &\leq n / 2^{i+1}.
    \end{align*}
    It follows that $a_{i+1} + q_{i+1} < n$. Consequently, either $a_{i+1} + q_{i+1} = (a_i + \lambda_0 q_i) + 2(a_{i+1} - (a_i + \lambda_0 q_i))$ or $a_{i+1} + q_{i+1} = a_i + (\lambda_0 + 1)(a_i+q_i-a_i)$. Hence, $a_{i+1} + q_{i+1}$ is in the forward closure of $\{a_0,\ldots,a_{i+1}\}$. Therefore $q_{i+1}$ has all desired properties.
\end{itemize}

\paragraph*{Conclusion}
For all $0 \leq i \leq k$, there exists $q_i \in \{1,\ldots,n - a_i\}$ such that
\begin{enumerate}
    \item  $q_i \leq n / 2^i$, and
    \item  $a_i + q_i = n$ or $a_i + q_i$ is in the forward closure of $\{a_0, \ldots,
    a_i\}$.
\end{enumerate}
\end{proof}

\begin{lemma}\label{lem:irreduciblebound}
Let $R(P) = \{0 = a_0 < a_1 < \ldots < a_k\}$ be the irreducible period set of a string of length $n$. Then $k \leq \log_2(n)$.
\end{lemma}
\begin{proof}
    It follows from the Lemma~\ref{lem:periodbound} that there exists an integer $q_k \in \{1, \ldots, n - a_k\}$ such that $n / 2^k \geq q_k$. 
    Hence $k \leq \log_2(n)$.
\end{proof}
To count the number of irreducible period sets, we count the number of possibilities for each $a_i$ with $1 \leq i \leq k$. We know that $a_0 = 0$ is fixed. The other $a_i$ take values in the set $\{1,\ldots,n-1\}$.
\begin{lemma}\label{lem:choicebound}
Let $0 \leq i \leq k - 1$. Then $a_{i+1}$ can take at most $2^{1-i}n - 1$ possible values given $a_0,\ldots,a_i$.
\end{lemma}
\begin{proof}
Let $q_i$ be defined as in Lemma~\ref{lem:periodbound}. We distinguish 3 cases:
\begin{enumerate}
    \item If $a_{i + 1} \leq a_i + q_i$, there are at most $q_i - 1 \leq n / 2^i - 1$ possible values for~$a_{i + 1}$ (note that $a_{i+1} \neq a_i + q_i$, because $a_{i+1}$ cannot be in the forward closure of $\{a_0,\ldots,a_i\}$, nor can it be equal to $n$).
    \item If $a_{i + 1} \geq n - q_i$, there are at most $q_i \leq n / 2^i$ possible values for $a_{i + 1}$.
    \item In the remaining case, $a_{i+1} \in [a_i + q_i + 1, n - q_i - 1]$.
\end{enumerate}
Let us first show that case 3 is impossible.
For the sake of contradiction, assume we are in case 3.  Since $a_i + q_i < n$, we know that $a_i + q_i$ is in the forward closure of $\{a_0,\ldots,a_i\}$ (by property 2 from Lemma~\ref{lem:periodbound}). Hence $q_i$ is a period of $u[a_i\dd n-1]$. Moreover $a_{i+1} - a_i$ is also a period of $u[a_i\dd n-1]$. By the Fine and Wilf theorem, it follows that 
\begin{enumerate}[(a)]
    \item either $n - a_i < q_i + (a_{i+1} - a_i) - \gcd(q_i,a_{i+1} - a_i)$
    \item or $\gcd(q_i,a_{i+1} - a_i)$ is a period of $u[a_i\dd n-1]$.
\end{enumerate}
We are not in subcase (a) since by hypothesis $a_{i + 1} \leq n - q_i - 1$. Suppose we are in subcase (b). Note that $a_i + \gcd(q_i,a_{i+1} - a_i) \leq a_i + q_i < a_{i+1}$ and that $a_{i+1}$ is in the forward propagation of $\{a_0,\ldots,a_i, a_i + \gcd(q_i,a_{i+1} - a_i)\}$. It follows that $a_{i+1}$ is not an irreducible period, which is a contradiction. Therefore both subcases (a) and (b) are impossible.

Summing over cases 1 and 2 (since case 3 is impossible), we conclude that, given $a_0,\ldots,a_i$, there are at most 
$$(n/2^i - 1) + n/2^i + 0 = 2^{1-i}n - 1$$ 
possibilities for $a_{i+1}$.
\end{proof} 
Note that the bound of Lemma~\ref{lem:choicebound} is not tight: indeed, there are $n - 1$ possible values for $a_1$, while the lemma gives an upper bound of $2n - 1$. However, this bound suffices to prove our asymptotic result. 
Since an autocorrelation is uniquely defined by its irreducible period set, it suffices to count the possible such sets $\{a_0, \ldots, a_k\}$ for all possible values of $k$. Recall that $a_0$ is fixed at 0 and that $k \leq \log_2(n)$ by Lemma~\ref{lem:irreduciblebound}. We thus derive a bound on the total number of autocorrelations by taking the product of all possibilities for $a_{i+1}$ with $i$ going from $0$ to $k - 1$ and sum this over all integers $k$ from $1$ to ${\lfloor\log_2(n)\rfloor}$,  as follows:
\begin{align*}
      \kappa_n =|\Gamma_n| = |\Lambda_n| &  \leq \sum_{k = 1}^{\lfloor\log_2(n)\rfloor}\prod_{i = 0}^{k-1} \left(2^{1-i}n - 1\right)\\
      &  \leq \sum_{k = 1}^{\lfloor\log_2(n)\rfloor}\left(\left(2^{2 - k}n - 1\right)\prod_{i = 0}^{k-2} 2^{1-i}n\right).\\
      \intertext{Writing $2^{2 - k}n\prod_{i = 0}^{k-2} 2^{1-i}n$ and $\prod_{i = 0}^{k-2} 2^{1-i}n$ in exponential form, we get}
      \kappa_n &\leq \sum_{k = 1}^{\lfloor\log_2(n)\rfloor}\Bigg(\exp\left(\frac{-k(k-3)\ln(2)}{2} + k\ln(n)\right)\\&\quad - \exp\left(\frac{-(k-1)(k-4)\ln(2)}{2} + (k-1)\ln(n)\right)\Bigg).\\
      \intertext{Observe that this is a telescoping sum, so all but two terms cancel out.}
      \kappa_n&\leq \exp\left(\frac{-\lfloor\log_2(n)\rfloor(\lfloor\log_2(n)\rfloor-3)\ln(2)}{2} + \lfloor\log_2(n)\rfloor\ln(n)\right) - 1\\
      \intertext{Since $\frac{d}{dk}\left(\frac{-k(k-3)\ln(2)}{2} + k\ln(n)\right) = \frac{(-2k+3)\ln(2)}{2} + \ln(n)$ is positive for all $k \leq \log_2(n)$, we have}
       \kappa_n&< \exp\left( \frac{\ln(n)(3\ln(2)-\ln(n))}{2\ln(2)} + \frac{\ln^2(n)}{\ln(2)}\right)\\
       &= \exp\left( \frac{3\ln(n)}{2} + \frac{\ln^2(n)}{2\ln(2)}\right).
      \end{align*}
Taking the natural logarithm of both sides and dividing by $\ln^2(n)$, we get that
$$  \frac{\ln(\kappa_n)}{\ln^2(n)}  \leq \frac{1}{2 \ln(2)} + \frac{3}{2\ln(n)}, $$
thereby proving Theorem~\ref{thm:upp}.

\end{proof}
\begin{corollary}[Asymptotic Convergence of $\kappa_n$]\label{thm:con} Let
$\kappa_n$ be the number of autocorrelations of length $n$. Then 
$$\frac{\ln\kappa_n}{\ln^2(n)} \rightarrow \frac{1}{2\ln(2)} \quad \text{as} \quad n \rightarrow \infty.$$
\end{corollary}

\begin{proof}
It follows from Theorem~\ref{thm:upp} that for $n \in \mathbb{N}_{\geq 2}$
$$\frac{\ln(\kappa_n)}{\ln^2(n)} \leq\frac{1}{2\ln(2)} + \frac{3}{2\ln(n)}  = \frac{1}{2\ln(2)} + o(1).$$
The lower bound for $\kappa_n$ from Theorem 5.1 in~\cite{Rivals} indicates that asymptotically 
\begin{align*}
  \frac{\ln(\kappa_n)}{\ln^2 (n)}
  & \geq \frac{1}{2\ln (2)}\left(1 - \frac{\ln(\ln(n))}{\ln(n)}\right)^2 + \frac{0.4139}{\ln(n)} - \frac{1.47123\ln(\ln(n))}{\ln^2 (n)} +O\left(\frac{1}{\ln^2 (n)}\right)
  \\
  & = \frac{1}{2\ln (2)} - O\left( \frac{\ln(\ln(n))}{\ln(n)} \right).
\end{align*}
Combining this lower bound with our upper bound, we obtain
$$ \frac{1}{2\ln(2)} - O\left(\frac{\ln\ln n}{\ln n}\right) \quad \leq \quad  \frac{\ln \kappa_n}{\ln^2(n)}  \quad \leq \quad  \frac{1}{2\ln(2)} + o(1).$$
Using the classic \emph{sandwich theorem}, we conclude that
$$\frac{\ln\kappa_n}{\ln^2(n)} \rightarrow \frac{1}{2\ln(2)} \quad \text{as}\quad n \to \infty$$
thereby proving the conjecture by Guibas and Odlyzko.
\end{proof}

\begin{figure}[htb]
    \centering
    \includegraphics[width=0.75\textwidth]{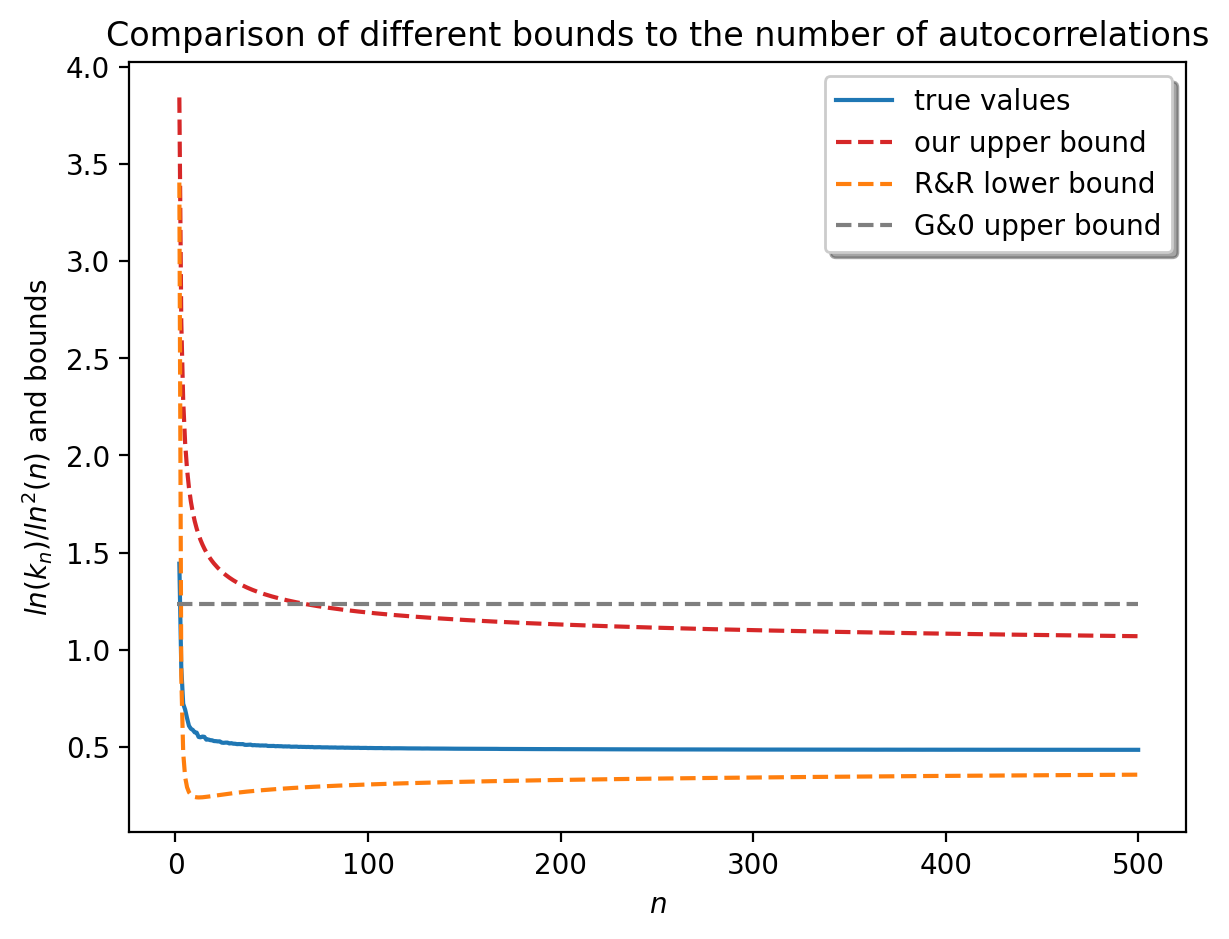}
    \caption{The true values of $\ln k_n/\ln^2(n)$ for $n \leq 500$ are compared to: the upper bound of Guibas \& Odlyzko (G\&O upper bound)~\cite{Guibas}, the Fröberg lower bound (R\&R lower bound)~\cite{Rivals}, and our upper bound. Our upper bound seems not so tight: the reason is that $n$ is small, as $\ln 500 \approx 6.2$.}  
    \label{fig:kn}
\end{figure}

The known values of \(\kappa_n\) are recorded in entry {A005434} (see \url{https://oeis.org/A005434}) of the On-Line Encyclopedia of Integer Sequences~\cite{oeis}. Because the enumeration of \(\Gamma_n\) takes exponential time, the list of \(\kappa_n\) values is limited to a few hundred. In Figure~\ref{fig:kn}, we compare the values of  \(\kappa_n\) with the so-called Fröberg lower bound from~\cite{Rivals}, the upper bound of Guibas and Odlyzko~\cite{Guibas}, and our new upper bound. The figure illustrates the improvement brought by the new upper bound compared to that given by Guibas and Odlyzko~\cite{Guibas}. At $n=500$, the lower bound, our new upper bound, and the values of \(\kappa_n\) clearly differ, meaning the sequences are far from convergence at $n=500$.

\section{Correlation}
In this section, we show that the number of correlations between two strings of length $n$ has the same asymptotic convergence behaviour as the number of autocorrelations of strings of length $n$. \\
In~\cite{GO81b}, Guibas and Odlyzko introduced the notion of \emph{correlation} of two strings: it encodes the offset of possible overlaps between these two strings.
In~\cite{Guibas}, the same authors investigate the self-overlaps of a string, which is then encoded in an \emph{autocorrelation}.
Before we start, let us define precisely the notion of correlation (which is illustrated in Table~\ref{tab:cor}).

\begin{definition}[Correlation]\label{def:correlation}
For every pair of strings $(u, v) \in \Sigma^n \times \Sigma^m$, the correlation of $u$ over $v$ is the vector $t \in \{0, 1\}^n$ such that for all $k \in \{0, \ldots, n - 1\}$
$$t[k] = \begin{cases}1 & \text{if } u[i] = v[j] \text{ for all } i \in \{0,\ldots,n-1\}, j \in \{0,\ldots,m-1\}\\&\text{with } i = j + k,\\
0 & \text{otherwise}
         \end{cases}$$
         
\end{definition}
Intuitively, we can find correlations as follows. For each index $i \in \{0, \ldots, n-1\}$ we write $v$ below $u$ starting under the $i$th character of $u$. Then the $i$th element of the correlation is 1, if all pairs of characters that are directly above each other match, and $0$ otherwise. See Table~\ref{tab:cor} for an example.

\begin{table}
\begin{center}
\begin{tabular}{>{\columncolor[gray]{.8}}cccccccccccc>{\columncolor[gray]{.8}}c} 
  \rowcolor[gray]{.8}
  pos.  & 0 & 1 & 2 & \color{blue}{3} & 4 & 5 & 6 & 7 & 8 & 9 & 10  &
  \\ 
  $u$  & \texttt{a} & \texttt{a} & \texttt{b} & \texttt{b} & \texttt{a} & \texttt{a} & - & - & - & - & -  &$t$ \\ \hline
  $v$  & \texttt{b} & \texttt{a} & \texttt{a} & \texttt{b} & \texttt{a} & \texttt{a} & - & - & - & - &  &  0
  \\ 
        & - & \texttt{b} & \texttt{a} & \texttt{a} & \texttt{b} & \texttt{a} & \texttt{a} & - & - & - & - &  0
  \\
        & - & - & \texttt{b} & \texttt{a} & \texttt{a} & \texttt{b} & \texttt{a} & \texttt{a} & - & - & - &  0
  \\
        & - & - & - & \color{blue}{\texttt{b}} & \color{blue}{\texttt{a}} & \color{blue}{\texttt{a}} & \color{blue}{\texttt{b}} & \color{blue}{\texttt{a}} & \color{blue}{\texttt{a}} & - & - &  \color{blue}{1}
  \\
        & - & - & - & - & \texttt{b} & \texttt{a} & \texttt{a} & \texttt{b} & \texttt{a} & \texttt{a} & - &  0
  \\
        & - & - & - & - & - & \texttt{b} & \texttt{a} & \texttt{a} & \texttt{b} & \texttt{a} & \texttt{a} &  0
  \\
  \hline
\end{tabular}
\end{center}
\caption{The correlation of word $u = \mathtt{aabbaa}$ over word $v = \mathtt{baabaa}$ (both of length $6$) is $t = 000100$. This table is organized as Table~\ref{tab:auto} -- see the corresponding caption for details.}\label{tab:cor}
\end{table}


Observe, that if $v \in \Sigma^m$ is longer than $u \in \Sigma^n$, then the correlation of $u$ over $v$ equals the correlation of $u$ over $v[0\dd n-1]$. Conversely, any binary vector $t \in \{0, 1\}^n$ is the correlation of $u = t \in \{0, 1\}^n$ over $v = 1 \in \{0, 1\}^1$. Therefore we will restrict ourselves to the interesting case where both strings have the same length. 

Let $\Delta_n$ be the set of all correlations between two strings of the same length $n$ and let $\delta_n$ be the cardinality of $\Delta_n$.
We can characterize $\Delta_n$ as follows.
\begin{lemma}\label{lem:corrchar}
The set of correlations of length $n$ is of the form $$\Delta_n = \left\{0^{(n-j)} s_j\ \mid \  s_j \in \Gamma_j,\ j\in[0,n] \right\},$$
where $\Gamma_j$ is the set of autocorrelations of length $j$.
\end{lemma}
\begin{proof}
Let $t = 0^{(n-j)} s_j$ with $s_j$ the autocorrelation of some string $w$ of length $j$ with $0 \leq j \leq n$. Without loss of generality, $w$ does not start with the letter $\mathtt{a}$. 
Let $u = \mathtt{a}^{(n-j)}w$ and $v = w\mathtt{b}^{(n-j)}$. Observe that the correlation of $u$ over $v$ is precisely $0^{(n-j)} s_j = t$. Therefore $$\left\{0^{(n-j)} s_j\ \mid \  s_j \in \Gamma_j,\ j\in[0,n] \right\} \subseteq \Delta_n.$$

Conversely, let $u, v \in \Sigma^n$ and let $t'$ be the correlation of $u$ over $v$. We can write $t'$ in the form $0^{(n-j)}s_j$, where $s_j$ is a binary string starting with $1$ (or is empty). If $s_j$ is the empty string, then it is the only autocorrelation of length $0$. Otherwise, there is a 1 at position $n-j$, which indicates that $u[n-j\dd n-1] = v[0\dd j-1]$. Moreover, $s_j$ is the correlation of $u[n-j\dd n-1]$ over $v$. It follows that $s_j$ is exactly the autocorrelation of $u[n-j\dd n-1] = v[0\dd j-1]$. Therefore
$$\Delta_n \subseteq \left\{0^{(n-j)} s_j\ \mid \  s_j \in \Gamma_j,\ j\in[0,n] \right\}.$$
\end{proof}


In the above characterization, we consider strings over a finite alphabet and found that a correlation depends on some autocorrelation. As it is known that $\Gamma_n$ is independent of the alphabet size (provided $|\Sigma| > 1$), the reader may wonder whether the number of correlations depends on it. In Appendix~\ref{sec:independence}, we show that the set of correlations for equally long strings is independent of the alphabet size,  provided that $\Sigma$ is not unary.

Now we have characterized $\Delta_n$, we can easily deduce its cardinality.
\begin{lemma}\label{lem:sum}
Let $\kappa_n$ be the number of autocorrelations of length $n$ and $\delta_n$ the number of correlations between two strings of length $n$. Then
$$\delta_n = \sum_{j = 0}^n \kappa_j.$$
\end{lemma}
\begin{proof}
Since autocorrelations do not start with a zero, no two strings of the form $0^{(n-j)} s_j$ with $s_j \in \Gamma_j$ and $j\in[0,n]$ are the same. Therefore
\begin{align*}\delta_n = |\Delta_n| &= \left|\left\{0^{(n-j)} s_j\ \mid \  s_j \in \Gamma_j,\ j\in[0,n] \right\}\right|
 = \sum_{j = 0}^n |\Gamma_j|
 = \sum_{j = 0}^n \kappa_j.\end{align*}
\end{proof}
\begin{theorem}[Asymptotic Convergence of $\delta_n$]\label{thm:asym2} Let $\delta_n$ be the number of correlations between two strings of length $n$. Then
$$\frac{\ln\delta_n}{ \ln^2(n)} \rightarrow \frac{1}{2\ln(2)} \quad \text{as} \quad n \rightarrow \infty.$$
\end{theorem}
 \begin{proof}
From Lemma~\ref{lem:choicebound} we know that for all $n \in \mathbb{N}_{\geq 2}$ 
$$\ln(\kappa_n)  \leq \frac{\ln^2(n)}{2 \ln(2)} + \frac{3\ln(n)}{2}.$$
It follows that for all $n \in \mathbb{N}_{\geq 2}$ we have
\begin{align*}\frac{\ln(\delta_n)}{\ln^2(n)} &= \ln \left(\sum_{i = 0}^n\kappa_n\right)/\ln^2(n)\\
&\leq \ln \left(2 + (n - 1)\exp\left(\frac{\ln^2(n)}{2 \ln(2)} + \frac{3\ln(n)}{2}\right)\right)/\ln^2(n)\\
&\leq \left(\frac{\ln^2(n)}{2 \ln(2)} + \frac{3\ln(n)}{2} + \ln(n)\right)/\ln^2(n)\\
&= \frac{1}{2 \ln(2)} + o(1) \quad \text{as} \quad n \rightarrow \infty.
\end{align*}
Conversely, using the fact that $\delta_n \geq \kappa_n$, we find $$\frac{\ln\delta_n}{ \ln^2(n)} \geq \frac{\ln\kappa_n}{ \ln^2(n)} =  \frac{1}{2\ln(2)} + o(1)\quad \text{as} \quad n \rightarrow \infty.$$
Again, by the sandwich theorem we conclude
$$\frac{\ln\delta_n}{ \ln^2(n)} \rightarrow  \frac{1}{2\ln(2)}\quad \text{as} \quad n \rightarrow \infty.$$\end{proof}


\bibliography{references.bib} 
\appendix

\section{Omitted proofs} \label{sec:omitted}
\begingroup
\def\thelemma{\ref{lem:multiply}}
\begin{lemma}
Let $p$ be a period of $u \in \Sigma^n$ and $k \in \mathbb{Z}_{\geq 0}$ such that $kp < n$. Then $kp$ is also a period of $u$.
\end{lemma}
\begin{proof}If $p = 0$ or $k = 0$, the statement trivially holds. Suppose $p \in \{1,\ldots, n-1\}$ and $k > 0$. If $i, j \in \{0, \ldots, n - 1\}$ such that $i \equiv j \mod kp$, then we also have $i \equiv j \mod p$, and hence $u[i] = u[j]$ by Definition~\ref{def:period1}. This shows $kp$ is a period of $u$ by Definition~\ref{def:period1}.
\end{proof}
\addtocounter{lemma}{-1}
\endgroup
\begingroup
\def\thelemma{\ref{lem:add}}
\begin{lemma}
Let $p$ be a period of $u \in \Sigma^n$ and $q$ a period of the suffix $w = u[p\dd n-1]$. Then $p + q$ is a period of $u$. Moreover, $p + kq$ is also a period of $u$ for all $k \in \mathbb{Z}_{\geq 0}$ with $p + kq < n$.
\end{lemma}
\begin{proof}
By Definition~\ref{def:period2} of period, the fact that $p$ is a period of $u$ implies $u[0\dd n-p-1] = u[p\dd n-1]$, while $q$ is a period of $w$ implies $w[0\dd n - p - q - 1] = w[q\dd n-p-1]$. As $w$ is the suffix of $u$ starting at position $p$, we can combine the above results to find that 
\begin{align*}
    u[0\dd n-p-q-1] &= u[p\dd n-q-1] = w[0\dd n-p-q-1]\\
    &= w[q\dd n-p-1] = u[p+q\dd n-1],
\end{align*}
which indicates that  $p + q$ is a period of $u$. Moreover, if $p + iq$ is a period of $u$ for some $i \in \mathbb{N}$, then we can similarly show that $p + (i+1)q$ is also a period of $u$ if $p + (i+1)q < n$. It follows by induction that $p + kq$ is a period of $u$ for all $k \in \mathbb{N}$ with $p + kq < n$. The case $k = 0$ is trivial.
\end{proof}
\addtocounter{lemma}{-1}
\endgroup
\begingroup
\def\thelemma{\ref{lem:subtract}}
\begin{lemma}
Let $p, q$ be periods of $u \in \Sigma^n$ with $0 \leq q \leq p$. Then the prefix and the suffix of length $n-q$ have the period $p-q$.
\end{lemma}
\begin{proof}
Since $p, q$ be periods of $u \in \Sigma^n$ with $0 \leq q \leq p$, we have
\begin{align*}
    u[0\dd n-p-1] &= u[p\dd n-1] && \text{(by periodicity $p$)}\\
    &= u[p-q\dd n-q-1] && \text{(by periodicity $q$).}
\end{align*}
It follows that $u[0\dd n - q - 1]$ has period $p - q$. Similarly the suffix of $u$ of length $(n - q)$  also has period $p - q$.
\end{proof}  
\addtocounter{lemma}{-1}
\endgroup
\begingroup
\def\thelemma{\ref{lem:divide}}
\begin{lemma}
Suppose $p$ is a period of $u \in \Sigma^n$ and there exists a substring $v$ of $u$ of length at least $p$ and with period $r$, where $r|p$. Then $r$ is also a period of $u$.
\end{lemma}
\begin{proof} If $p = 0$, then $r=0$ and the lemma trivially holds. 

Otherwise $p$ is non-zero. Let $i,j \in [0,n-1]$ with $i \equiv j \mod r$. We can write $v = u[h \dd  k]$ with $0 \leq h < k \leq n-1$. Since $v$ has length at least $p$, there exist $i', j' \in [h,k]$ such that $i \equiv i'\mod p$ and $j \equiv j' \mod p$. By Definition~\ref{def:period1} of period, we have $u[i] = u[i']$ and $u[j] = u[j']$. Note that $i' \equiv i \equiv j \equiv j' \mod r$, because $r \mid p$. Applying 
Definition~\ref{def:period1} again, we obtain $u[i'] = u[j']$. It follows that $u[i] = u[i'] = u[j'] = u[j]$. Therefore $r$ is a period of $u$.\end{proof}
\addtocounter{lemma}{-1}
\endgroup

\section{Independence of alphabet} \label{sec:independence}
Guibas and Odlyzko showed that for every autocorrelation, there exists a string over a binary alphabet with that autocorrelation~\cite{Guibas}. A nice alternative constructive proof appears in~\cite{HALAVA2000298}.  We will now show that the same holds for arbitrary correlations of equally long strings.
\begin{corollary}
For any $t \in \Delta_n$, there exist $u, v \in \{\mathtt{a}, \mathtt{b}\}^n$ such that the correlation of $u$ over $v$ is $t$.
\end{corollary}
\begin{proof}
Let $t$ be the correlation of $u'$ over $v'$ with $u', v' \in \Sigma^n$. By Lemma~\ref{lem:corrchar}, we can write $t = 0^{(n-j)} s_j$, where $s_j \in \{0,1\}^j$ is the autocorrelation of $u'[n-j\dd n-1] = v'[0\dd j-1]$. By the result of Guibas and Odlyzko, we know that there also exists some binary string $w \in \{\mathtt{a}, \mathtt{b}\}^j$ with the same autocorrelation. Without loss of generality, we can assume that $w$ starts with $\mathtt{b}$. It follows that the constructed strings $u = \mathtt{a}^{(n-j)}w$ and $v = w\mathtt{b}^{(n-j)}$, which have a correlation of $t$ by the proof of Lemma~\ref{lem:corrchar}, use the same binary alphabet. 
\end{proof}
We conclude that the number of correlations between strings of equal length is alphabet-independent (i.e. every alphabet of size at least 2 gives rise to the same set of correlations). 
\begin{remark} Such a binary string $w$ can be constructed from $u'[n-j\dd n-1]$ in linear time using the algorithm of Halava, Harju and Ilie~\cite{HALAVA2000298}. Therefore $u$ and $v$ can also be constructed in linear time given $u'$ and $v'$.
\end{remark}

\end{document}